\documentclass[aip,amsmath,amssymb, preprint]{revtex4-2}
\usepackage{graphicx}
\usepackage{dcolumn}
\usepackage{bm}
\usepackage[utf8]{inputenc}
\usepackage[T1]{fontenc}
\usepackage{mathptmx}
\usepackage{calc}
\usepackage{color}
\graphicspath{{figures/}}
\newlength{\depthofsumsign}
\begin{document}
\preprint{AIP/123-QED}
\title{Exciton Transfer in Organic Photovoltaic Cells: A Role of Local and Nonlocal Electron-Phonon Interactions in a Donor Domain}
\author{Mauro Cainelli}
\email{cainelli.mauro@gmail.com}
\affiliation{Department of Chemistry, Graduate School of Science, Kyoto University, Kyoto 606-8502, Japan}

\author{Yoshitaka Tanimura}
\email{tanimura.yoshitaka.5w@kyoto-u.jp}
\affiliation{Department of Chemistry, Graduate School of Science, Kyoto University, Kyoto 606-8502, Japan}

\begin{abstract}
We theoretically investigate an exciton transfer process in a donor domain of organic photovoltaic cells focusing on the roles of local and nonlocal electron-phonon interactions.  Our model consists of a three-level system described by the Holstein-Peierls Hamiltonian coupled to multiple heat baths for local and nonlocal molecular modes characterized by Brownian spectral distribution functions. We chose tetracene as a reference donor molecule, where the spectral distribution functions of the local and nonlocal modes exist. We then employ the reduced hierarchy equations of motion (HEOM) approach to simulating the dynamics of the system under the influence of the environment as a function of the electron-phonon coupling strength and temperature. We rigorously calculate the reduced density matrix elements to explain the timescale of dynamics under the influence of the dissipative local and nonlocal modes. The results indicate that the strong nonlocal electron-phonon interaction under  high temperature conditions favors the exciton transfer process and enhances the efficiency of organic photovoltaic materials, while the lifetime of the exciton becomes shorter due to a low frequency local mode.
\end{abstract}

\date\today
\maketitle

\section{Introduction}
\label{sec:level1}
The development of better photovoltaic devices is a matter of great interest in the investigation of renewable energy sources for a wide variety of systems including biology- and material-oriented venues. Organic solar cells are promising materials because of its low fabrication cost, lightweight and the flexibility of design in comparison to silicon-based materials exist in the market.\cite{ref1,ref2} Typical composition consists of polymers of small molecules as donors and fullerene derivatives as acceptors, although non-fullerene derivatives have also exhibited promising results.\cite{ref:38} However, the power conversion efficiency of organic solar cells remains with maximum values of 11-17\% that are still much lower than the inorganic counterparts.\cite{ref:3,ref:4} Moreover, the mechanism on the organic devices is more complicated, mainly because of the strong electron-phonon interaction due to the low dielectric constant, where the quantum coherence between the electron and phonon plays an essential role. Thus, for the investigation to improve the efficiency of such systems, we have to employ fully quantum mechanical descriptions of broad validity.\cite{ref:6}

During the process, sunlight is absorbed predominantly by the molecular donor with the creation of strongly bounded electron-hole pairs (excitons). The excitons then diffuse to the heterojunction, in which they dissociate into electrons and holes, and are then collected at the corresponding electrodes.\cite{ref:9}
While most of the electronic dynamics occur in the active layer, recent studies indicate that timescales and length scales of these processes depend on the local geometry and electrostatic field. \cite{ref:13,ref:14} Therefore, for the design of photovoltaic devices, understanding and controlling the morphology is significant.\cite{ref:10,ref:11,ref:12}  
Although charge transport and charge separation processes in organic solar cells have been studied intensively, investigations concerning the role of incoherent (hopping) \cite{ref:15} and coherent (delocalization) \cite{ref:16,ref:17,ref:18} transfer mechanisms in different structure and physical conditions have not been well explored.\cite{ref:19,ref:20,ref:21} 

As recent theoretical analysis indicated, the effect of the nonlocal (intermolecular) electron-phonon coupling on charge transport processes, in addition to that of the local (intramolecular) one, is essential to account for the energy transfer mechanism on such devices.\cite{ref:22} In heterojunction blends, the entire process is ultrafast and takes place within 100fs \cite{ref:43,ref:44} due to the presence of short lived delocalized excited species,\cite{ref:45,ref:46,ref:47} which associate with the formation of hot charge transfer (CT) states.\cite{ref:49} The hot CT states are assumed to be the main contribution to the generated photocurrent, because they allow us to create a charge separation before the thermalization process occur, while incoherent charge separation mechanisms are characterized by slower separation of charges via multiple hopping steps between localized states. In organic solar cells, both mechanisms are assumed to be present.

It should be noted that the predominant dynamical processes in solar cells are a system dependent, in which the molecules used as donor and acceptor, geometry disposition, and electron-phonon coupling strengths determine the power conversion efficiency. In order to account for the problem of this kind, processes such as light absorption, energy- and charge-transports in organic semiconductors must be taken into account by simulating dynamics in both electronic and molecular degrees of freedom simultaneously.\cite{ref:51} A widely used approach in order to properly treat vibrational modes in the framework of quantum mechanics is based on the Holstein-Peierls model,\cite{ref:26} where the Holstein (local) part of the Hamiltonian describes the variation of the site energies\cite{ref:23} and the Peierls (nonlocal) part of it represents the modulation of the transfer integrals.\cite{ref:24} Although the importance of nonlocal electron-phonon coupling on the charge transport properties has been realized recently,\cite{ref:25} the Hamiltonian of this model is not diagonalizable and the analysis of the system is not easy.  Thus, the effects of the local and nonlocal couplings are usually treated collectively under the mean field polaron (MFP) approximation,\cite{ref:27} ignoring the interaction between the local and nonlocal modes. Moreover, because the electronic processes occur in condensed phases, where the surrounding molecules provide the thermal fluctuation and dissipation, we have to adapt open quantum dynamics treatment for the investigation of time-irreversible reaction dynamics.\cite{ref:28,ref:29} 

In order to study electron transfer problems in connection to open quantum dynamics theory, a commonly used model considers the electronic states coupled to an intermediate harmonic oscillator, which is further coupled to a heat bath.\cite{ref:31} Then, the harmonic mode can be included in the bath by carrying out a canonical transformation, which leads to a multilevel system coupled to the heat bath with the Brownian spectral distribution (BSD) function.\cite{ref:32} Such a system can be treated using the HEOM formalism, in a numerically rigorous manner,\cite{ref:60} even in the low temperature case.\cite{ref:34, ref:35, ref:36} The HEOM formalism can handle not only the strong system-bath coupling but also quantum coherence between the system and bath that is important for the investigation of the organic materials, as demonstrated in photosynthesis antenna systems\cite{ref:63,ref:64,ref:65,ref:66} and DNA.\cite{ref:67}
Furthermore, this method does not require any approximation, most notably the rotating wave approximation (RWA) or the MFP approximation. 

The HEOM approach has been applied to the case of solid state materials described by a deformation potential \cite{ref:33} and the Holstein Hamiltonian.\cite{ref:61} In the Holstein case, because we can study only a small system with the finite number of phonon modes associated with the system site,  special treatment is necessary to maintain the stability of the equations.\cite{ref:62} The situation is the same in the present case, and thus we further introduce the heat-bath into the regular Holstein-Peierls model.

In this paper, we present a study of the nonlocal electron-phonon coupling effect and the relation with a local coupling effect on the charge transference process in organic photovoltaic devices, particularly in the donor layer. We employ the HEOM approach for a BSD function to simulate the non-Markovian dynamics of the Holstein-Peierls model that is further coupled to the heat-bath.

The organization of this paper is as follows: In Sec. ~\ref{sec:Holstein}, we introduce the Holstein-Peierls + bath (HPB) model. Then, the HEOM for the HPB model is explained. In Sec.~\ref{sec:result}
 we present the details of the calculation and the results of the simulation for various conditions. Sec.~ \ref{sec:Conclusion}
 is devoted to concluding remarks. 

\section{The HEOM for a Holstein-Peierls + bath system}
\label{sec:Holstein}

The Holstein-Peierls model employs a tight-binding picture of an electronic system that is linearly coupled to an optical phonon mode at each site. In the present study, we further introduce a harmonic oscillator heat-bath.

The total Hamiltonian is expressed in terms of the electron Hamiltonian ($\hat{H}_{el}$), phonon Hamiltonian ($ \hat{H}_{ph}$) and electron-phonon interaction with the heat-bath ($\hat{H}_{ph-B}$) as
\begin{eqnarray}
\hat{H}_{tot} = \hat{H}_{el} + \hat{H}_{el-ph} +\hat{H}_{ph-B}, 
\label{H_tot}
\end{eqnarray}
where
\begin{eqnarray}
\hat{H}_{el} = \sum_i \varepsilon_i^0\hat{a}_i^{\dagger}\hat{a}_i + \sum_{i\ne j}t_{ij}^0\hat{a}_i^{\dagger}\hat{a}_j,
\end{eqnarray}
\begin{eqnarray}
\hat{H}_{el-ph} &=& \sum_{i}\sum_{ {\alpha} }  {\hat V}_{\alpha}^{i} \left(\hat{b}_{\alpha} ^{\dagger}+\hat{b}_{\alpha} \right) + \sum_{ij}\sum_{ {\alpha} }  {\hat V}_{\alpha}^{ij} \left(\hat{b}_{\alpha} ^{\dagger}+\hat{b}_{\alpha} \right)
\end{eqnarray}
with 
\begin{eqnarray}
 {\hat V}_{\alpha}^{i} = \frac{1}{\sqrt{{\cal N}_C}}\hbar\Omega_{\alpha}  g_i^{\alpha}\hat{a}_i^{\dagger}\hat{a}_i, \nonumber \\
{\hat V}_{\alpha}^{ij} = \frac{1}{\sqrt{{\cal N}_C}}\hbar\Omega_{\alpha} 
 g_{ij}^{\alpha}\hat{a}_i^{\dagger}\hat{a}_j ,
\end{eqnarray}
and
\begin{eqnarray}
\hat{H}_{ph-B} &=& \sum_{\alpha} \hbar\Omega_{\alpha} \left(\hat{b}_{\alpha} ^{\dagger}\hat{b}_{\alpha} +\frac{1}{2}\right) +  \sum_{\alpha}\sum_{k=1}^{N_{\alpha}} \hbar\omega_k \left(\hat{b}_k^{\dagger}\hat{b}_k+\frac{1}{2}\right) \nonumber \\ 
                          &   &+ \sum_{\alpha} \left(\hat{b}_{\alpha}^{\dagger}+\hat{b}_{\alpha} \right) \sum_{k=1}^{N_{\alpha}}  c_k^{\alpha}\left(\hat{b}_k^{\dagger}+ \hat{b}_k \right). 
\end{eqnarray}
Here, $\hat{a}_i^\dagger$ ($\hat{a}_i$) is the creation (annihilation) operator of the electronic quasi-particles (electrons, holes, or Frenkel excitons) and
$\varepsilon_i^0$  and $t_{ij}^0$ are the on-site electronic energy and the ampliture of the transfer integral for the $i-j$th electronic states that correspond to the diagonal and off-diagonal elements of the on-site Hamiltonian, respectively. The creation (annihilation) operator of the phonon  (or vibron) mode ${\alpha}$  with the frequency $\Omega_{\alpha} $ is expressed as \ $\hat{b}_{\alpha} ^\dagger$ ($\hat{b}_{\alpha} $) .  The local and nonlocal system-bath interactions are expressed as ${\hat V}_{\alpha}^{i}$ and ${\hat V}_{\alpha}^{ij}$ with  the dimensionless local and nonlocal electron-phonon couplings strength $g_i^\alpha$ and $g_{ij}^\alpha$, respectively.  The constant ${\cal N}_C$ is the number of unitary cells considered. The operator  $\hat{b}_k^\dagger$ ($\hat{b}_k$) is the creation (annihilation) operator of the heat-bath that can be characterized by the oscillator-bath coupling strength and the inverse temperature $\beta \equiv 1/k_{\mathrm{B}}T$, where $k_\mathrm{B}$ is the Boltzmann constant. The bath system is typically modeled by the spectral distribution function (SDF), defined by
\begin{eqnarray}
  J_{\alpha} (\omega) \equiv \sum_{\alpha}\sum_{k=1}^{N_{\alpha}} \left(c_{k}^{\alpha}\right)^2 \delta(\omega-\omega_{k}).
  \label{eq:J_wgeneral}
\end{eqnarray}
For the heat bath to be an unlimited heat source possessing an infinite heat capacity, the number of heat bath oscillators $ N_{\alpha}$ is effectively made infinitely large by replacing $J_\alpha(\omega)$ with a continuous distribution.
This Hamiltonian can be regarded as either empirical or first-principles model, whether the system parameters are obtained from experimental data or quantum chemistry calculations.
The electron-phonon interactions produce a time-dependent variation of the transport parameters and thus introduce a dynamic disorder: Local and nonlocal electron-phonon couplings correspond to diagonal and off-diagonal dynamic disorder mechanisms.

The challenges in using the Holstein-Peierls model are the following.\cite{ref:5} First, the Hamiltonian is not exactly solvable. Second, the commonly used approximations to solve this Hamiltonian are not fully satisfied for the investigation of organic semiconductors. Third, it is difficult to find efficient and reliable schemes to evaluate the system parameters ($\varepsilon_i^0$, $t_{ij}^0$, $\omega_{\alpha} $, $g_i^{\alpha} $,and  $g_{ij}^{\alpha} $).

In general, the second point in the above arises as a consequence of the first and, in order to study the model, several approximations have been introduced: First, the electron-electron interaction is treated in a framework of mean field theory. Second, the amplitude $t_{ij}^0$ must be small in comparison with the excitation energy of the intramolecular mode. Third, the electron-phonon coupling is considered to be a linear form. Four, the nonlocal coupling is treated under the band renormalization theory, assuming the collective effect of low-frequency modes (MFP approximation). Five, phonon modes are treated under rigid-body approximation that ignores the internal molecular vibrations with environmental degrees of freedom that consist of the intermolecular translational and librational molecular modes, in addition to the acoustic modes.
Moreover, in order to understand the structural effects on the charge dynamics and the role of mixing morphology on the efficient pathway to free charge photogeneration, it is necessary to account for the individual effects of the couplings at different phonon frequencies and the interaction between intra- and intermolecular vibrational modes.

In order to overcome some of the limitation mentioned above, here we employ the HEOM formalism. The MFP approximation can be avoided because the Hamiltonian does not require to be diagonalized. Thus, the phonon modes are treated dynamically employing the system-bath model and large values of $t_{ij}^0$ can be treated by choosing an appropriate hierarchy depth. Although in this work we limited our analysis to the linear electron-phonon couplings case, the extension to study nonlinear coupling is also possible. 

By using the canonical transformation, the Hamiltonian in Eq.\eqref{H_tot} can be expressed in terms of the creation and annihilation operators of the heat-bath + primary oscillation ($\hat{b}_k'^{\dagger}$ and $\hat{b}_k'$, respectively) in the following general form
\begin{eqnarray}
\hat{H}_{tot} &=& \hat{H}_{el} +\sum_{\alpha}\sum_{k=1}^{N_{\alpha}} \sum_{i} d_k^{ii\alpha} \hat{a}_i^{\dagger}\hat{a}_i(\hat{b}_k'^{\dagger}+ \hat{b}_k' ) \nonumber \\
                       &   &+ \sum_{\alpha}\sum_{k=1}^{N_{\alpha}}\sum_{ij}f_k^{ij\alpha}\hat{a}_i^{\dagger}\hat{a}_j (\hat{b}_k'^{\dagger}+ \hat{b}_k' )+ \sum_{\alpha}\sum_{k=1}^{N_{\alpha}}\hbar\omega_k \left(\hat{b}_k'^{\dagger}\hat{b}_k'+\frac{1}{2}\right) \label{H_tot2},
\end{eqnarray}
where $d_k^{ii\alpha}$ and $f_k^{ij\alpha}$ are the dimensionless constants that represent the local and nonlocal interactions of the system with the oscillator-bath.
Assuming that the SDF of the later to be $J_\alpha(\omega)=\gamma_{\alpha}\omega$,  we have the BSD defined as
\begin{eqnarray}
J_{\alpha}'(\omega) = \frac{\hbar\lambda_{\alpha}}{2\pi} \frac{\gamma_{\alpha}\Omega_{\alpha}^2\omega}{(\Omega_{\alpha}^2-\omega^2)^2+\gamma^2\omega^2},
\end{eqnarray}
where  $\lambda_\alpha$ relates to the reorganization energy and nonlocal relaxation energy which accounts for the displacement of the excited state in a relation to the ground state in coordinate space as a consequence of the local and nonlocal electron-phonon interactions, respectively. The parameter $\gamma_\alpha$ is the coupling strength between the oscillator and the bath, which is related to the peak width of $ J_{\alpha}'(\omega)$.

The reduced hierarchy equations of motion for the Brownian distribution can be expressed as (see Appendix A)
\begin{widetext}
\begin{eqnarray}
	&\frac{\partial} {\partial t}&\hat{\rho}_{\{n^{\alpha},m^{\alpha}; j_1^{\alpha}...j_{K^{\alpha}}^{\alpha}\}} (t) 
\nonumber \\ 
& = & 
-\left[\frac{i}{\hbar} \hat{H}_{el}^{\times} + \sum_{\alpha} \left\{ \frac{(n^{\alpha}+m^{\alpha})}{2} \gamma_{\alpha}- i(n^{\alpha}-m^{\alpha})\zeta_{\alpha} + \sum_{k=1}^{K^{\alpha}} j_k^{\alpha} \nu_k^{\alpha} - \hat{\Xi}_{\alpha} \right\} \right] \hat{\rho}_{\{n^{\alpha},m^{\alpha}; j_1^{\alpha}...j_{K^{\alpha}}^{\alpha}\}} (t)\nonumber \\ 
										 &     & +\sum_{\alpha} \hat{V}^{\times}_{\alpha} \bigg[\hat{\rho}_{\{n^{\alpha}+1,m^{\alpha}; j_1^{\alpha}...j_{K^{\alpha}}^{\alpha}\}} (t)+ \hat{\rho}_{\{n^{\alpha},m^{\alpha}+1; j_1^{\alpha}...j_{K^{\alpha}}^{\alpha}\}} (t) \bigg] \nonumber \\
& &
 + \sum_{\alpha} n^{\alpha}\hat{\Theta}_-^{\alpha} \hat{\rho}_{\{n^{\alpha}-1,m^{\alpha}; j_1^{\alpha}...j_{K^{\alpha}}^{\alpha}\}} (t) +\sum_{\alpha} m^{\alpha}\hat{\Theta}_+^{\alpha} \hat{\rho}_{\{n^{\alpha},m^{\alpha}-1; j_1^{\alpha}...j_{K^{\alpha}}^{\alpha}\}} (t) \nonumber \\
 										 &     & + \sum_{\alpha}\sum_{k=1}^{K^{\alpha}} \hat{V}_{\alpha}^{\times} \hat{\rho}_{\{n^{\alpha},m^{\alpha}; j_1^{\alpha}...j_{k+1}^{\alpha},...j_{K^{\alpha}}^{\alpha}\}} (t) + \sum_{\alpha}\sum_{k=1}^{K^{\alpha}} j_k^{\alpha} \nu_k^{\alpha} \hat{\Psi}_k^{\alpha}  \hat{\rho}_{\{n^{\alpha},m^{\alpha}; j_1^{\alpha}...j_{k-1}^{\alpha},...j_{K^{\alpha}}^{\alpha}\}} (t),\label{eq:heom_HPB}
\end{eqnarray}
\end{widetext}
where the first term in brackets is the Liouvillian ($\hat{\mathcal{L}}^{(n,m)}$) that involves the temperature correction terms.
For a high number of hierarchy elements, the term $(n^\alpha+m^\alpha) \gamma_\alpha/2 + \sum_{k=1}^{K^\alpha} j_k^\alpha \nu_k^\alpha$  becomes much larger than the characteristic time of the system. In this case, the hierarchy can be truncated by the terminator.\cite{ref:30}

\section{Results and Discussion}
\label{sec:result}
\begin{figure}[!t]
        \center{\includegraphics[scale = 0.7]
        {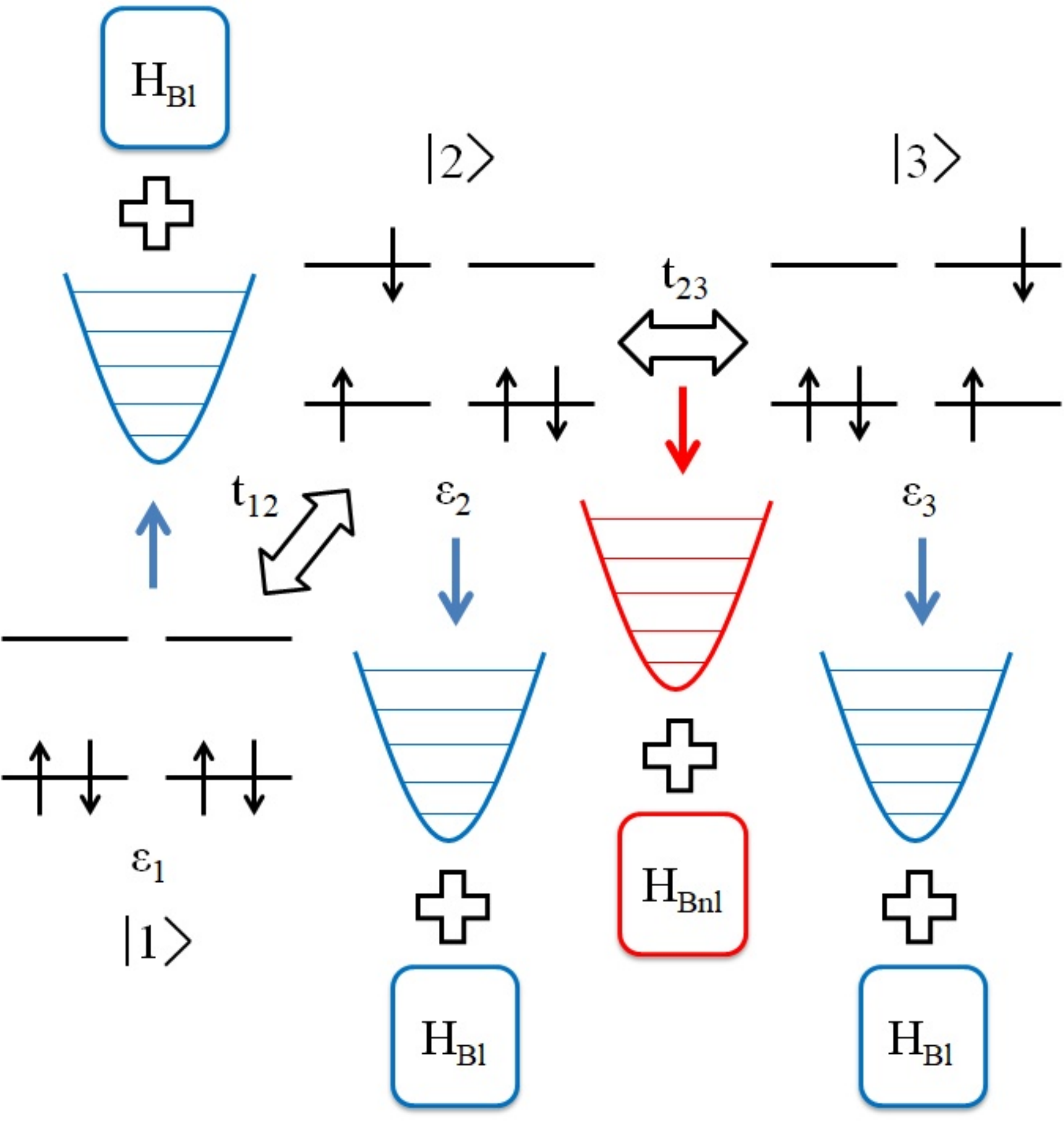}}
        \caption{\label{fig1} Schematic view of the model that is described by the Hamiltonian given in Eqs.(1)-(5). The ground state and the two excited states are coupled to a local mode that is further coupled to the local bath ($H_{Bl}$), respectively. The excited states are coupled to the nonlocal mode that is further coupled to the nonlocal bath ($H_{Bnl}$).}
\end{figure}
\begin{figure}[!h]
	\center{\includegraphics[scale = 0.7,keepaspectratio]
	{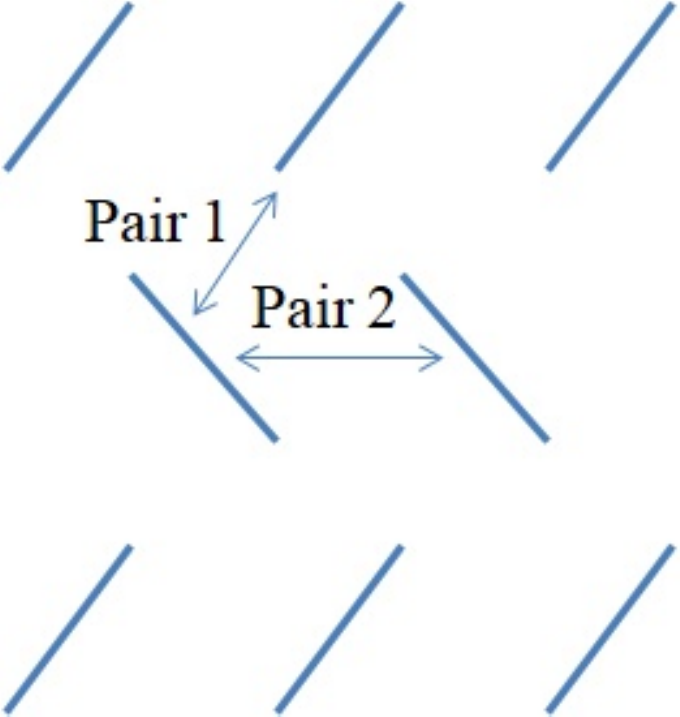}}
	\caption{\label{fig2}Two configurations of the excited state pathways. Each pair represents the dimer molecule illustrated in Fig.~\ref{fig1}.}
\end{figure}

We chose tetracene as a reference donor molecule under study where the SDFs of the electron-phonon coupling have been known.\cite{ref:37} In order to simulate the exciton transfer process between the molecules, we consider a multi-state system in which each state represents a dimer in the donor layer. While we might simulate a molecular chain in the framework of the present model by increasing the number of excited states, here we consider a simple three-state system in order to investigate the role of local and nonlocal modes.  Thus, the system is described by a single ground state and two excited states expressed as $\big | 1 \big\rangle $, $\big | 2 \big\rangle $, and $\big | 3 \big\rangle$, respectively. As illustrated in Fig.~\ref{fig1}, the excited states $\big | 2 \big\rangle $ and $\big | 3 \big\rangle$ interact with the nonlocal and local modes, while the ground state $\big | 1 \big\rangle $ only interacts with the local mode. The local and nonlocal modes are further coupled to their own heat-bath. We then consider the transfer between the two excited states in two directions, represented by two different pair of geometry dispositions of the dimer (Fig.~\ref{fig2}), whose difference is described by the values of the transfer integral and the nonlocal coupling strength. 

The exciton  Hamiltonian is then expressed in the matrix form as
\begin{gather}
\hat{H}_{el} = \hbar \begin{bmatrix} \omega_1 & \Delta_{12} & 0 \\  \Delta_{21} & \omega_2 & \Delta_{23} \\  0 & \Delta_{32} & \omega_3  \end{bmatrix},
\end{gather}
where $\hbar\omega_i$ is the Frenkel exciton energy of the $i$th state ($\varepsilon_i$) and $\hbar\Delta_{ij}$ is the coupling strength between the $i$th and $j$th states (transfer integral $t_{ij}$).
For all of our computations, we fixed the eigenenergy of the system as $\omega_1=0$cm$^{-1}$ and $\omega_2=\omega_3=500$cm$^{-1}$. We then use the excitation energy as the characteristic frequency of the system $\omega=500$cm$^{-1}$ with the time scale $1/{\omega}=66.71$fs. Although the band gap between the ground and singlet state in oligoacenes is of the order of 2-2.5eV ($16000-20000$cm$^{-1}$), here we chose a much smaller value for $\omega$ to qualitatively investigate the effect of phonon interactions upon ultrafast dynamics with suppressing numerical costs. The off-diagonal elements representing the interaction between the ground and excited state are $\Delta_{12}=\Delta_{21}=0.32\omega$. It should be noted that, while similar single crystals, which include rubrene\cite{ref:73,ref:74} and pentacene\cite{ref:69} have been modeled and investigated, the parameter values of the off-diagonal element of tetracene has not been well explored. Here we employ the value of pentacene as a representative case, while the interactions between the excited states are chosen to be $\Delta_{23}=\Delta_{32}=1.23\omega$ for the pair 1 and $\Delta_{23}=\Delta_{32}=0.12\omega$ for the pair 2, respectively. We consider the room temperature case ($T=300$K) with the value of $\beta\hbar\omega=0.382$ and the low temperature case (T=10K) with the value of $\beta\hbar\omega=11.45$. 

The system-bath interactions for the local and nonlocal modes are defined in the matrix form as
\begin{gather}
\hat{V}^{i} = \begin{bmatrix} -0.5 & 0 & 0 \\  0 & 0.5 & 0 \\  0 & 0 & 0.5  \end{bmatrix},
\end{gather}
and
\begin{gather}
\hat{V}^{ij} = \begin{bmatrix} 0 & 0 & 0 \\  0 & 0 & 1 \\  0 & 1 & 0  \end{bmatrix}.
\end{gather}
The diagonal and off-diagonal elements of the system-bath coupling relate to the local and nonlocal dynamic disorder mechanisms. Although the SDFs of local and nonlocal baths consist of two or three oscillator modes, here we consider one primary mode to simplify the model and analysis. Then, the coupling strength between the local/nonlocal oscillator and the bath, which acts as the inverse noise correlation time of the BSD, is fixed to $\gamma=0.05\omega$. The initial condition of the system at $t=0$ is set by $\rho_{22}(0)=1$ and $\rho_{11}(0)=\rho_{33}(0)=0$. We simulate the time evolution of the reduced density matrix by numerically integrating the HEOM with respect to the time 
with a time step of ${0.001}/{\omega}$, using the fourth-order Runge-Kutta method. The depth and the truncation number of hierarchy are chosen to $N=65$ with $K=0$ for the high temperature case and $N=12$ with $K=3$ for the low temperature case, respectively.

\begin{figure}[!t]
        \center{\includegraphics[scale = 0.5,keepaspectratio]
        {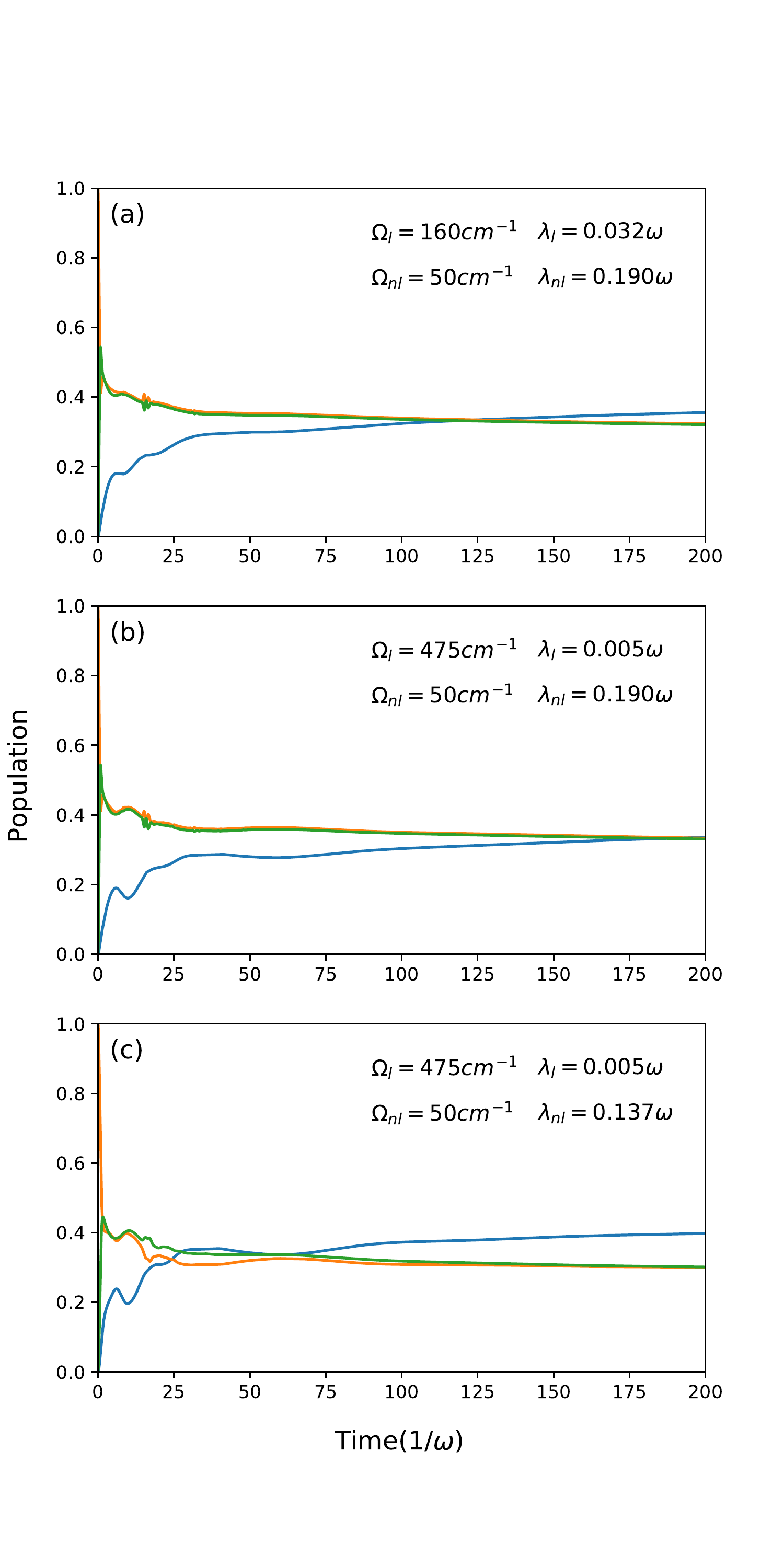}}
        \caption{\label{fig3} Time-evolution of the reduced density matrix elements $\rho_{11}(t)$ (blue), $\rho_{22}(t)$ (orange), and $\rho_{33}(t)$ (green) at $T=300K$ for different parameter values of local and nonlocal oscillator bath. Here, (a) and (b) represent the pair 1 case and (c) represents the pair 2 case. The very early stage of the time evolution in each figure is presented in Figs.~\ref{figB1}(a)-(c) in Appendix B.}
\end{figure}

The time-evolution of the reduced density matrix for different sets of bath modes at room temperature is shown in Fig.~\ref{fig3}. For the local/nonlocal baths, we consider (a) $(\Omega_{l}^1,~\lambda_{l}^1)=(160,~ 0.032\omega)$/$(\Omega_{nl}^1,~\lambda_{nl}^1)=(50,~0.19\omega)$ and (b) $(\Omega_{l}^1,~\lambda_{l}^1)=(475,~0.005\omega)$/$(\Omega_{nl}^1,~\lambda_{nl}^1)=(50,~0.19\omega)$ for the pair 1, while (c) $(\Omega_{l}^2,~\lambda_{l}^2)=(475,~0.005\omega)$/$(\Omega_{nl}^2,~\lambda_{nl}^2)=(50,~0.137\omega)$ for the pair 2, with the unit of cm$^{-1}$. 

As illustrated in Fig.~\ref{fig3}(a), the local mode of lower frequency slightly enhances the exciton relaxation at a longer time period in comparison to that of higher frequency presented in Fig.~\ref{fig3}(b), because the coupling strength of the local mode $\lambda_{l}$ is larger in the former case. The population relaxation in Fig.~\ref{fig3}(b) is also slower than that in Fig.~\ref{fig3}(c), because, in the case of pair 1, the exciton transfer between $\big | 2 \big\rangle $ and $\big | 3 \big\rangle $ is favored due to the large nonlocal electron-phonon coupling $\lambda_{nl}$ and transfer integral (excited state interaction) $\Delta_{23}$.  The very early stage of the time-evolution in each figure is presented in Figs. \ref{figB1} (a)-(c) in Appendix B, respectively, and it is shown that this transfer is characterized by an ultrafast coherent oscillation in the time period of $t$=1 [1/$\omega$] for pair 1 (Figs.~\ref{figB1}(a) and ~\ref{figB1}(b)) and $t$=1.5 [1/$\omega$] for pair 2 (Fig.~\ref{figB1}(c)).
These facts indicate that the local and nonlocal modes play an opposite role: The nonlocal mode promotes ultrafast exciton transfer as a function of the coupling strength of the nonlocal mode, $\lambda_{nl}$, that suppresses the exciton relaxation to the ground sate, whereas the local mode enhances the relaxation through the localization of the exciton as a function of the coupling strength of the local mode, $\lambda_{l}$. 

It should be mentioned that the small oscillatory motion of the excited state population observed at approximately $t$=20 [1/$\omega$] in Figs.~\ref{fig3}(a) and ~\ref{fig3}(b) are numerical errors that arise from an insufficient depth of hierarchy elements $N$ due to the computational limitation of the HEOM calculation. This can be suppressed by increasing $N$, although the calculations become computationally more expensive. 

\begin{figure}[!t]
        \center{\includegraphics[scale = 0.5,keepaspectratio]
        {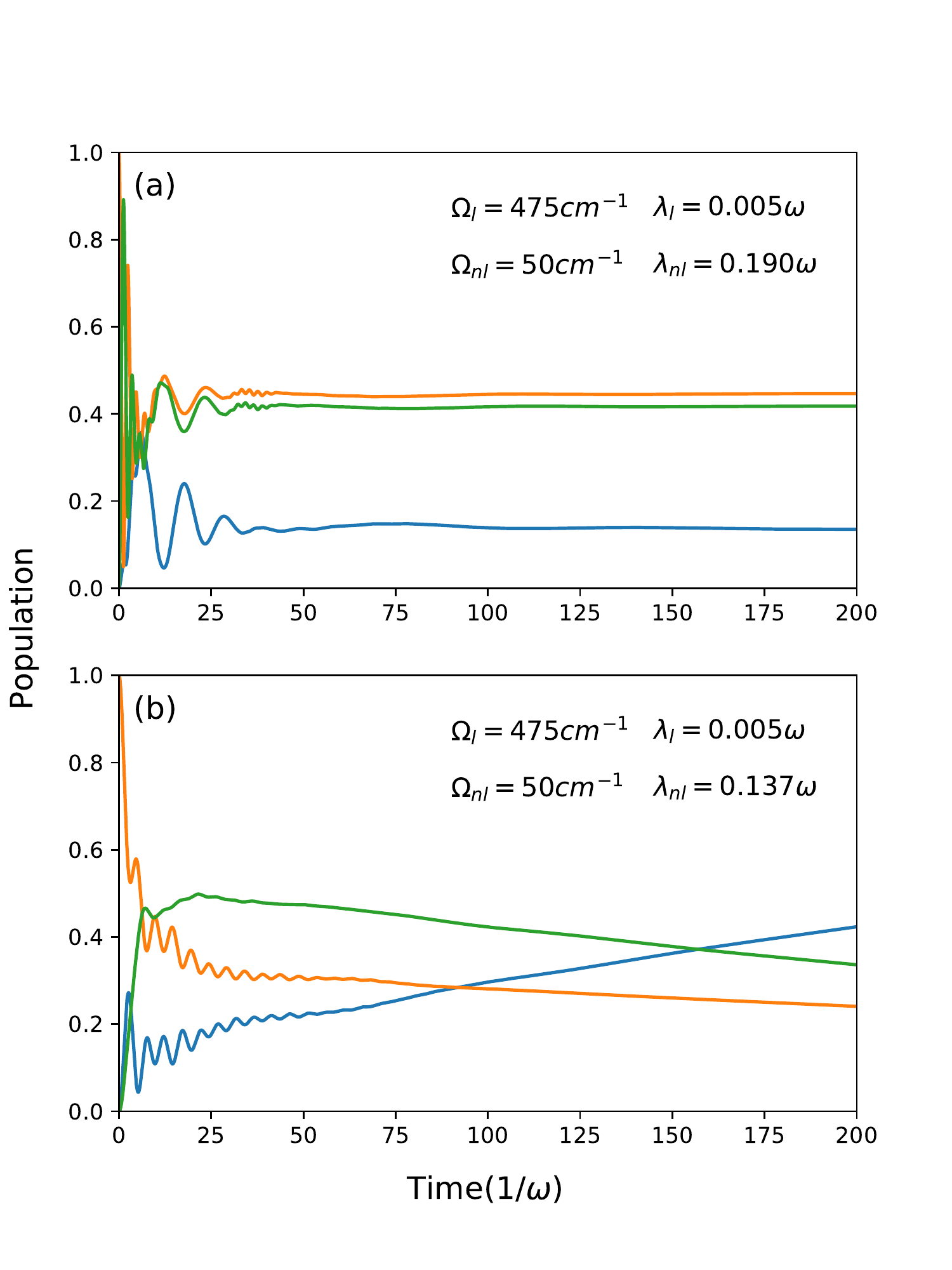}}
        \caption{\label{fig4} Time evolution of the reduced density matrix elements $\rho_{11}(t)$ (blue), $\rho_{22}(t)$ (orange), and $\rho_{33}(t)$ (green)  at $T=10K$ for (a) pair 1 and (b) pair 2 with the same set of the bath parameters presented in Figs.~\ref{fig3}(b) and \ref{fig3}(c), respectively. The very early stage of the time-evolution in each figure is presented in Figs.~\ref{figB2}(a) and  \ref{figB2}(b) in Appendix B.}
\end{figure}

Next we discuss the low temperature cases ($T$=10K) of  Figs.~\ref{fig3}(b) and ~\ref{fig3}(c) in Figs.~\ref{fig4}(a) and \ref{fig4}(b), respectively. The initial stage of Figs.~\ref{fig4}(a) and \ref{fig4}(b) are separately plotted in Figs.~\ref{figB2}(a) and ~\ref{figB2}(b) in Appendix B, respectively. We observe ultrafast coherent oscillations between the excited states in the time period of $t$=1.2 [1/$\omega$] for the pair 1 and $t$=7.5 [1/$\omega$] for the pair 2, respectively. The frequency of the ultrafast oscillation in the low temperature case is slower than that in the room temperature case, and thus the population exchange between $\big | 2 \big\rangle$ and $\big | 3 \big\rangle$ toward the equilibrium state takes longer time. In fact, we observe that, in the case of the pair 2, the exciton relaxation starts at time $t$=2.5 [1/$\omega$], which is shorter than the time period of the nonlocal oscillation. This indicates that, under the low temperature condition, the effect of the bath through the local and nonlocal couplings is suppressed, while the system dynamics is determined from the interplay of the ground and the excited states coupling $\Delta_{12}$ and the excitation coupling $\Delta_{23}$. For the pair 1 with larger $\Delta_{23}$, the exciton transfer between $\big | 2 \big\rangle$ and $\big | 3 \big\rangle$ is favored and the relaxation to the ground state $\big | 1 \big\rangle$ becomes slower than that for the pair 2.

In this regards, although the ground state population of the thermal equilibrium state becomes larger for longer time period at lower temperature, the equilibration process that arises from the local and nonlocal baths is faster in the high temperature case.In any case, when the coupling strength between the excited states and the nonlocal coupling strength are larger, the relaxation to the ground state become slower. Thus, large nonlocal electron-phonon coupling seems to favors the mechanism of the exciton transfer among the excited states even in the weak excitation coupling strength condition, while the lifetime of the exciton is suppressed by low frequency local modes. Because the configuration of pair 1 satisfies both the strong excitation coupling and strong nonlocal coupling conditions, it is more favorable than that of pair 2.

Although it was found from investigations of polaron transport process that the nonlocal electron-phonon coupling promotes not only delocalization, but also localization due to a bandwidth narrowing effect, in particular for the strong coupling case at high temperatures,\cite{ref:70,ref:71,ref:72} the model system employed in this work is too small and further study in a larger system needs to be done for the verification.

\section{Conclusion}
\label{sec:Conclusion}
In this work, we introduced Holstein-Peierls + bath Hamiltonian in order to investigate the effects of local and nonlocal electron-phonon interactions for the exciton dynamics of organic molecules in photovoltaic cells. We employed the HEOM formalism to simulate the dynamics of excitons described by this model under various physical conditions for local and nonlocal environmental modes without employing approximations, most notably the RWA and MFP approximations. It should be noted that the situations that we considered here are classified as non-perturbative and non-Markovian system-bath interaction case in open quantum dynamics theory.

The calculated results demonstrate the effects of dynamic disorder on the order of the reported time scale of the exciton processes that are characterized by the presence of ultrafast coherent oscillations, in particular at low temperatures. We found that a stronger nonlocal electron-phonon coupling favors the exciton transfer between excited states and may be a key to enhancing the efficiency of exciton transfer processes even in materials with low excitation couplings. Moreover, the effect seems to be favored at high temperature, while the number of low frequency local bath modes should be suppressed.

In the present work, we limited our analysis to a small system with a specific parameter set focusing on a role of local and nonlocal system-phonon interactions. Because the exciton transfer is a long range effect, an extension of the present investigation to a larger system should be realized to provide better insight in this topic, specially regarding further discussions between the delocalization/localization mechanisms. We plan to investigate this direction in a future study.

\begin{acknowledgments}
    The financial support from The Kyoto University Foundation is acknowledged. M.C. is supported by the Japanese Government (MEXT) Postgraduate Scholarships. 
\end{acknowledgments}
    
\section*{Data Availability}
The data that support the findings of this study are available from the corresponding author upon reasonable request.

\appendix
\section{\label{sec:level1}Reduced Hierarchy Equations of Motion for Holstein-Peierls Model}

We consider the density matrix of the total system with the factorized initial condition. The reduced density matrix element is then expressed in the path integral form as
\begin{eqnarray}
\rho(z,z',t) &=& \iint_{z(t)=z}^{z'(t)=z'}\mathcal{D}[z,z']e^{iS_0[z,z']/\hbar}\nonumber \\
	        &   & \times \rho(z_0,z'_0,t_0)e^{-iS_0[z,z']/\hbar}\mathcal{F}[z,z',t],
\end{eqnarray}
where $\iint_{z(t)=z}^{z'(t)=z'}\mathcal{D}[z,z']$ represents the functional integral of a set of Grassmann variables, which describe the system states, $\rho(z_0,z'_0,t_0)$ is the factorized system part density, $S_0$ is the action for the system Hamiltonian (Lagrangian), and $\mathcal{F}[z,z',t]$  is the called Feynman-Vernon influence functional that includes all the information regarding the bath and the system bath-interaction.\cite{ref:52,ref:53, ref:59,ref:68}

Considering the time dependent kernels $iL_1(t)=i\int_{0}^{\infty}d\omega J(\omega)\sin(\omega t)$ and $L_2(t)=\int_{0}^{\infty}d\omega J(\omega)\coth(\beta\hbar\omega/2)\cos(\omega t)$ that correspond to the fluctuation and dissipation effects, respectively, introduced by the bath,\cite{ref:52,ref:54,ref:55} the influence functional can be written, evaluating the contour integrals for the Brownian spectral distribution as\cite{ref:35, ref:36,ref:57,ref:58}
\begin{eqnarray}
	\mathcal{F}[z,z',t] & = &exp\Bigg\{-\int_{t_0}^{t} ds \int_{t_0}^{s} du V^{\times}(s)\nonumber \\
	                     	         &     &\times \bigg[\Theta_{-}(u)e^{-\big(\frac{\gamma}{2}-i\zeta\big)(s-u)}+\Theta_{+}(u)e^{-\big(\frac{\gamma}{2}+i\zeta\big)(s-u)}\bigg]\Bigg\}\nonumber \\
			         &     &\times \exp\Bigg[\int_{t_0}^{t} ds \int_{t_0}^{s} du V^{\times}(s) \sum_{k=1}^{\infty}\Psi_k(u)\nu_k e^{-\nu_k(s-u)}\Bigg]\nonumber \\
		                    &     &\times \exp\bigg[\int_{t_0}^{t}ds\Xi(s)\bigg],
\end{eqnarray}
where
\begin{eqnarray}
\Theta_{\pm}=\frac{\lambda\omega_0^2}{2\zeta}\Bigg\{\pm\coth\bigg[\frac{\beta\hbar}{2}\Big(i\frac{\gamma}{2}\mp\zeta\Big)\bigg]V^{\times}(u)\mp V^{\circ}(u)\Bigg\},
\end{eqnarray}
\begin{eqnarray}
\Psi_k(u)=\frac{4\lambda\gamma\omega_0^2}{\beta\hbar}\frac{\nu_k}{(\omega_0^2+\nu_k^2)^2-\gamma^2\nu_k^2}V^{\times}(u),
\end{eqnarray}
\begin{eqnarray}
\Xi(s) = V^{\times}(s)\sum_{k=K+1}^{\infty}\Psi_k(s).
\end{eqnarray}
Here, $\zeta=\sqrt{\omega_0^2-\gamma^2/4}$ for $\gamma<2\omega_0$ and $\nu_k=(2\pi/\beta h)k$ are the Matsubara frequencies, $\beta={1}/{k_B T}$ is the inverse temperature, and $k_B$ is the Boltzmann constant.

The commutator and anticommutator of the interaction operator $V$ are expressed as $V^{\times}(t)=V(z(t))-V(z'(t))$ and $V^{\circ}(t)=V(z(t))+V(z'(t))$, and the electron-phonon couplings are introduced in the spectral distribution. The local and nonlocal transition rates are determined from the diagonal and off-diagonal elements of $V$, respectively.

Thanks to the exponential function form of the kernels, any higher-order time derivative of the correlation functions is also expressed in terms of exponential functions. These terms are called the auxiliary density matrices (ADMs) in the HEOM formalism and are necessary to describe the non-Markovian effects of the environment. The HEOM for the Brownian distribution can be derived by calculating the time derivative on the left and right side of the factorized system density and the influence functional.\cite{ref:57,ref:58} Summing over all bath modes results in Eq.~\eqref{eq:heom_HPB}.

For a high number of hierarchy elements, the hierarchy can be truncated by the terminator
\begin{eqnarray}
\frac{\partial}{\partial t} \hat{\rho}_{j_1…j_k}^{(n,m)} (t) = -\bigg[\frac{i}{\hbar} \hat{H}_{el}^{\times}- i(n-m) \zeta - \hat{\Xi}\bigg] \hat{\rho}_{j_1…j_k}^{(n,m)} (t).
\end{eqnarray}
The total number of hierarchy elements can be evaluated as $L=(N+M+K+1)/(K+1)!(N+M)!$ with the total number of terminator elements $L_{term}=(N+M+K)/K!(N+M)!$, where $N$ is the hierarchy depth for  $n, m_+$ and $m_-$, and $M$ is the increase of the hierarchy for each spectral mode ($M=2$ for $m_+$ and $m_-$ in the case of the Brownian oscillator). In the HEOM formalism, each member of the hierarchy is coupled to the lower and higher terms, and the value for the initial density element is the exact solution of the total Hamiltonian defined by $\rho_{0…0}^{(0,0)} (t)$, which includes all the system-bath interactions.

\section{\label{sec:level2}The ultrafast oscillations of the density matrix elements in the very early stage}
\begin{figure}[!h]
        \center{\includegraphics[scale = 0.5,keepaspectratio]
        {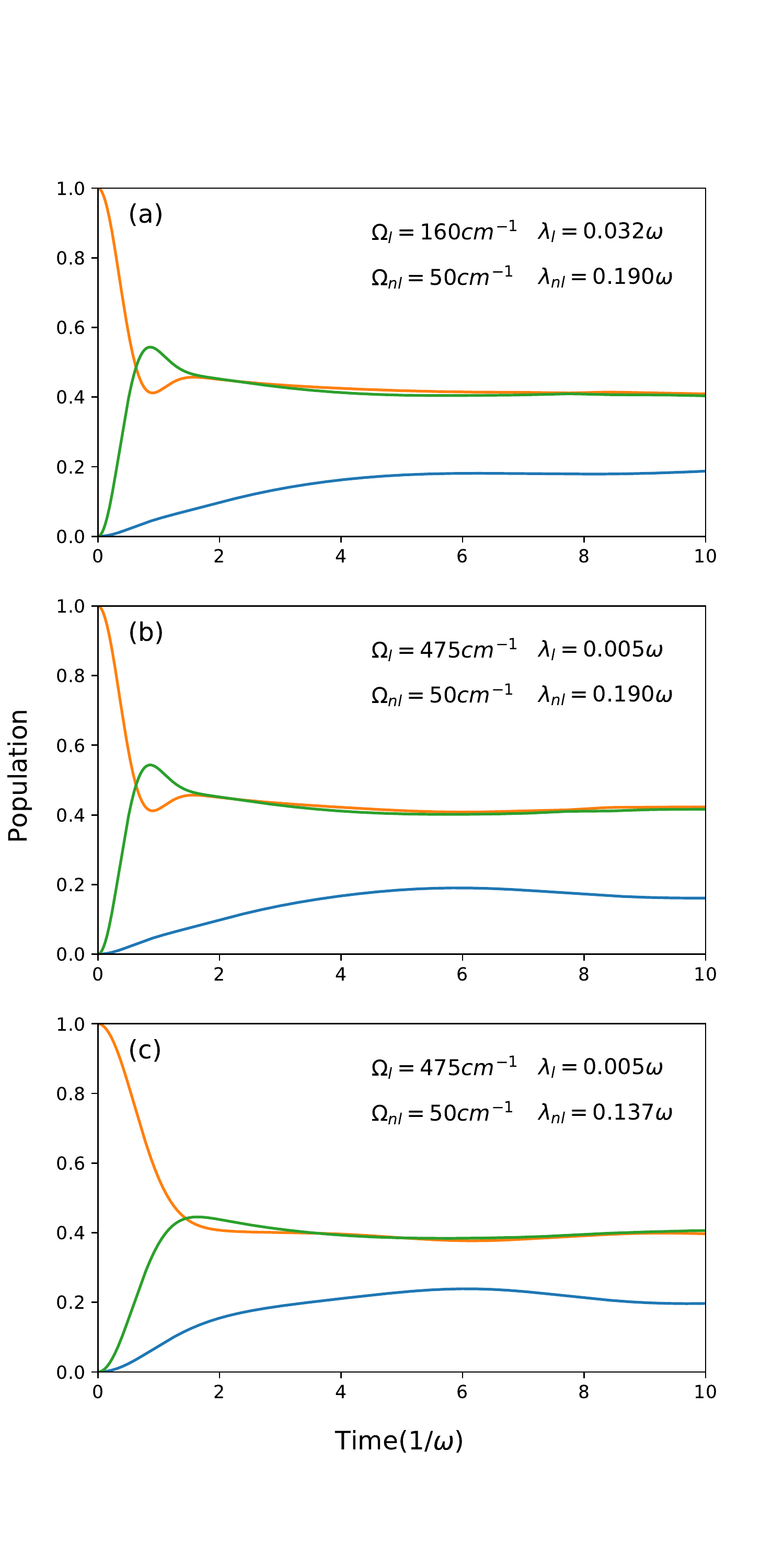}}
        \caption{\label{figB1}Very early stage of the time-evolution of the reduced density matrix elements $\rho_{11}(t)$ (blue), $\rho_{22}(t)$ (orange), and $\rho_{33}(t)$ (green) at $T=300$K presented in Figs.~\ref{fig3} (a)-(c) as Figs. (a)-(c), respectively, are illustrated.  Here, (a) and (b) correspond to the case of pair 1 and (c) corresponds to the case of pair 2.}
\end{figure}

\begin{figure}[!h]
        \center{\includegraphics[scale = 0.5,keepaspectratio]
        {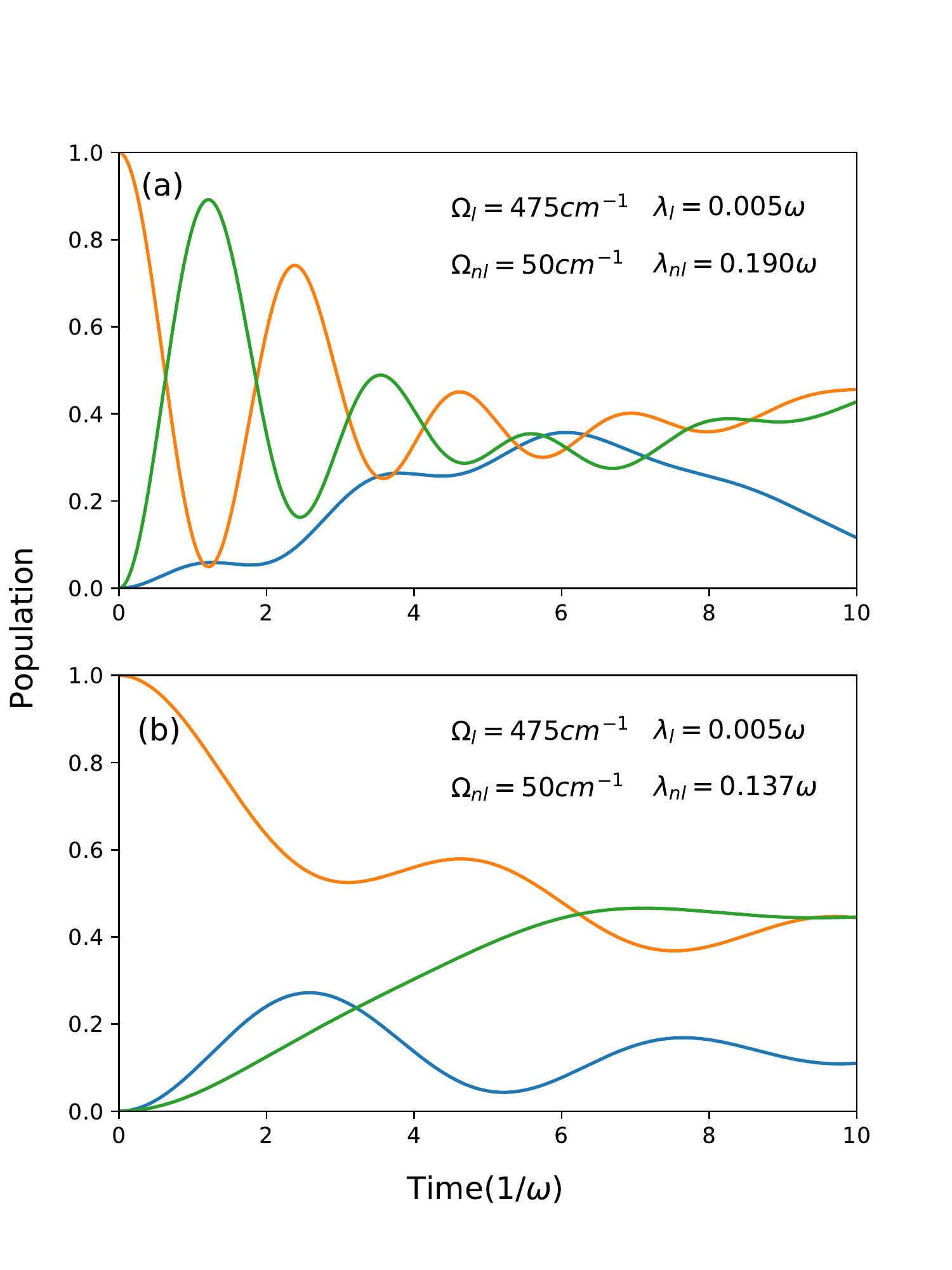}}
        \caption{\label{figB2} Very early stage of the time-evolution of the reduced density matrix elements $\rho_{11}(t)$ (blue), $\rho_{22}(t)$ (orange), and $\rho_{33}(t)$ (green) at $T=10$K presented in Figs.~\ref{fig4} (a) and (b) as Figs. (a) and (b), respectively, are illustrated.}
\end{figure}

Here, we replot the very early stage of the time-evolution ($t \le 10$ [1/$\omega$] of the reduced density matrix elements presented in Figs.~\ref{fig3} and ~\ref{fig4}, respectively. In each figure, we observe ultrafast coherent oscillations between $\big | 2 \big\rangle $ and $\big | 3 \big\rangle$. As it is illustrated in Figs.~\ref{figB1} (b) and ~\ref{figB1} (c) and Figs.~\ref{figB2} (a) and ~\ref{figB2} (b), the ultrafast oscillation is enhanced for larger $\lambda_{nl}$, which indicates that nonlocal electron-phonon interaction that relates to the dynamics disorder enhances the transition between the excited states. 

\clearpage

\bibliography{references1}
\end{document}